**The limits of the seismogenic zone in the epicentral region of the 26 Dec. 2004 Great Sumatra-Andaman earthquake: results from a seismic refraction and wide-angle reflection surveys and thermal modeling**

Klingelhoefer, F., Gutscher, M.-A., Ladage, S., Dessa, J.-X., Graindorge, D., Franke, D.,
André, C., Permana, H., Yudistira, T., Chauhan, A.

## Abstract

The 26th December 2004 Sumatra earthquake (Mw=9.1) initiated around 30 km depth and ruptured 1300 km of the Indo-Australian/Sunda plate boundary. During the Sumatra-OBS survey a wide-angle seismic profile was acquired across the epicentral region. A seismic velocity model was obtained from combined travel time tomography and forward modeling. Together with reflection seismic data from the SeaCause II cruise the deep structure of the source region of the great earthquake is revealed. Four to five km of sediments overlie the oceanic crust at the trench and the subducting slab can be imaged down to a depth of 35 km. We find a crystalline backstop 120 km from the trench axis, below the fore-arc basin. A high velocity zone at the lower landward limit of the ray-covered domain, at 22 km depth, marks a shallow continental Moho, 170 km from the trench.

The deep structure obtained from the seismic data was used to construct a thermal model of the forearc in order to predict the limits of the seismogenic zone along the plate boundary fault. Assuming 100-150° C as its updip limit, the seismogenic zone is predicted to begin 5-30 km from the trench. The downdip limit of the 2004 rupture as inferred from aftershocks is within the 350-450° C temperature range, but this limit is 210-250 km from the trench axis and is much deeper than the forearc Moho. The deeper part of the rupture occurred along the contact between the mantle wedge and downgoing plate.

## 1 Introduction

Subduction megathrusts produce by far the largest earthquakes on earth (Ruff and Kanamori, 1972; Stein and Okal, 2005). These large ruptures also cause significant vertical motions that can generate devastating tsunamis (Savage, 1983; Satake, 1993; Johnson et al., 1996; Fuji and Satake, 2007). Together these phenomena pose a substantial threat to populations and infrastructure located in coastal regions, such as most of the Pacific rim. The

30    surface area of the earthquake fault plane, together with the amount of slip, controls the magnitude of the earthquake (Wells and Coppersmith, 1994). The portion of the fault plane,

which ruptures, is said to be "seismogenic" and is bounded by an up-dip and a down-dip limit

(Tichelaar and Ruff, 1993; Oleskevich et al., 1999). The position of the updip limit exerts a

strong control on tsunami generation and the location of the downdip limit which commonly

35    lies close to the coast, influences the intensity of ground shaking here. Therefore, it is of

critical importance for the estimation and mitigation of these natural hazards to have reliable

estimates of these updip and downdip limits. However, there is still no consensus on the

physical processes which control these limits, and thus large uncertainties still exist for most

zones which have not experienced a great earthquake in recent times, e.g. Cascadia (Hyndman

40    and Wang, 1995; Khazaradze et al., 1999; Stanley and Villasenor, 2000). This work focuses

on the SE end of the rupture zone of the great Sumatra-Andaman earthquake of 26 Dec. 2004,

where the event initiated. We present new deep-penetration seismic data to constrain the

geometry and structure of the crust and upper mantle. Additionally, numerical modeling of

fore-arc thermal structure is performed in order to calculate the thermally expected limits of

45    the seismogenic zone. These results are compared to the observed distribution of aftershocks and to published source-rupture models in order to determine where the earthquake rupture initiated. The implications for the control of seismogenic rupture by the lithologies and temperatures along the plate contact are discussed.

## 1.1 The NW Sumatra margin and the M9.3 2004 earthquake

50    One of the first deep seismic studies on the Sumatran subduction zone consisted of a

two ship survey in 1966/67 (Curray et al., 1977). Their preferred interpretation of a profile off

central Java shows minor thickening of the oceanic crust at the trench and slightly seawards

which they propose to be caused by faulting. Later studies were carried out by Kiekhefer, et

al., 1980 using free floating sonobuoys, airguns and explosives on 5 margin-parallel lines

55    located offshore Nias and in the Nias basin. Based on the analysis of this dataset they propose a shallow Moho depth beneath the continental crust of only about 20 km along this segment of the Sumatra forearc.

During the SUMENTA I and II cruises in 1992 over 5000 km of 6 channel seismic data were acquired offshore Sumatra (Izart et al., 1994), which helped to constrain the

60 existence of the Mentawai micro-plate formed by partitioning of the oblique convergence of the subduction and strike-slip deformation of the upper plate (Malod and Kemal, 1996).

More recent multichannel and wide-angle reflection seismic studies of the SE-Sumatra to Java portion of the Sunda margin were carried during the GINCO project with the German research vessel Sonne in 1998 and 1999. Based on these data two main phases of fore-arc

65 basin and accretionary prism evolution were identified (Schlüter et al, 2002). Analysis of the wide-angle data provided images of the subducting plate to a depth of 20 km and additionally the crustal structure was modeled to a depth of 30 km using gravity data (Kopp et al., 2001). Their preferred model includes a shallow Moho at only 15 km depth beneath the Java margin (Kopp et al., 2002).

70 The 26th December 2004 earthquake (Mw = 9.1-9.3) is among the four largest earthquakes ever recorded and the largest of the last 40 years (Lay et al., 2005; Stein and Okal, 2005). The earthquake initiated off the NW Sumatra margin in the vicinity of Simeulue Island, and ruptured northwards past the Nicobar and Andaman islands, along 1300 km of the Indo-Australian/Sunda plate boundary (Lay et al., 2005). Source time studies (Ammon et al.

75 2005) as well as geodetic (Vigny et al., 2005) and tsunami inversions (Fuji and Satake, 2007), indicate that the rupture zone was widest (up to 200 km), co-seismic slip was largest (locally over 20m) and thus seismic moment release was greatest, in the region directly west of the NW tip of Sumatra, near Banda Aceh (Figure 1).

Several marine geophysical surveys were conducted after the earthquake on the

80 Sumatran margin. During the first cruise onboard the British military vessel HMS Scott multibeam bathymetry data were acquired on the southern part of the 2004 earthquake rupture zone (Henstock et al., 2006). From February - March 2005 a Japanese ocean-bottom seismometer array was deployed to record the numerous aftershocks of the earthquake. The authors proposed the existence of two splay faults from the distribution of the earthquakes

85 (Araki et al., 2006). During the French Aftershocks cruise, 20 OBS's were deployed in the region of largest slip offshore north-west of Sumatra. Examination of these data revealed the existence of two post-seismic active thrust faults and two other thrust faults, which were not post-seismically active (Sibuet et al., 2007). On the basis of the multi-beam bathymetry data of the French and British cruises, it was proposed that the upper plate deformation is strongly

90    influenced by the structure of the lower plate, which is characterized by North-South trending lineaments from the fossil Wharton spreading center (Graindorge et al., 2008).

Joint modeling of wide-angle and reflection seismic data from four profiles located

offshore Simeulue acquired by the R/V *Sonne* in 2006 shows a significant deepening of the

top of the oceanic crust towards the SE (Franke et al., 2008). The authors interpret, that a

95    ridge currently undergoing subduction in this region is responsible for the segmentation of the margin.

The Sumatra-OBS geophysical survey was conducted in July/Aug. 2006 onboard the

R/V Marion Dufresne in this area. The objective was to determine the crustal structure in the

epicentral region offshore NW Sumatra and in the zone of maximum co-seismic

100   displacement, some 300 km northward. We report here on the results of wide-angle seismic

modeling of Profile Sumatra-AB, a transect across the epicentral region of the 26 Dec. 2004

earthquake where the rupture initiated (Figure 2). Initial results from this survey are also

reported by (Dessa et al., 2009). Further results from interpretation of reflection seismic

profiles from a complementary cruise have been published recently (Singh et al., 2008). The

105   authors interpret some deep reflections west of Simeulue Island as representing broken slices

of oceanic crust, related to an oceanic mantle megathrust. They propose, that very strong

coupling, appropriate for brittle failure of mantle rocks, accounts for the occurrence of the

tsunamogenic event. Modeling of wide-angle  and reflection seismic data on a profile north of

Sumatra shows the presence of an active backthrust that is imaged down to at least 15 km

110   depth and reaches the surface near the western edge of the Aceh forearc basin (Chauhan et al.,

submitted to Geophys. J. Int.)  The authors suggest that co-seismic slip along the forearc backthrust boundary might provide an explanation for the generation of the tsunami.

Modeling was performed in combination with the pre-stack depth migration of multi-channel seismic data from the R/V Sonne cruise located close to the Sumatra-AB profile.

115   Modeling of the limits of the seismogenic zone was performed using a geometry taken from both the multi-channel and wide-angle seismic data.

## 1.2 The limits of the seismogenic zone

Several decades of research have been devoted to understanding how the physical and mechanical properties of the rocks vary along the plate boundary fault plane and thus to try to answer the question "why are earthquakes generated here?" (Byrne et al., 1988; Tichelaar and Ruff, 1993; Pacheco et al., 1993 Oleskevich et al., 1999). A comprehensive review of current knowledge on the limits of the seismogenic zone is given by Hyndman (2007). It has been suggested that the uppermost portion of the plate boundary, below the accretionary wedge behaves aseismically primarily due to the presence of high porosity, fluid rich sediments (Byrne et al., 1988). Whereas some margins have no accretionary wedge, in all cases it seems likely that the weakening effects of high fluid pressure within sediments along the plate interface can have an impact on seismogenesis (Hubbert and Rubey, 1959; Davis et al., 1983; Byrne and Fisher, 1990; Moore, et al., 1995; Saffer and Beakins, 2002). Other studies on the updip limit proposed a possible control due to the rheological behavior of clay minerals and sediments as these undergo low-grade metamorphic reactions and diagenesis. In particular the opal to quartz and smectite to illite and chlorite transitions occur primarily over a temperature range of 100-150°C with the higher grade minerals exhibiting a stick-slip rheology consistent with seismogenic behavior (Vrolijk, 1990; Moore and Saffer, 2001). However, recent laboratory studies are inconsistent at least for the experimental range of conditions tested, with the smectite - illite transition being the sole control on the updip limit of the seismogenic zone (Marone and Saffer, 2007).

There is also discussion on the mechanisms controlling the downdip limit. The two propositions are a thermal control due to the transition to dislocation creep at high temperatures (Scholz, 1990; Hyndman and Wang, 1993) or alternatively, that the presence of a highly serpentinized mantle wedge, which permits aseismic sliding, may define the downdip limit of seismogenic behavior (Hyndman et al., 1997). A global analysis of subduction megathrust earthquakes indicates that on average, the downdip limit occurs at a depth of 40 ± 5 km (Tichelaar and Ruff, 1993) though some subduction zones may have shallower downdip limits of about 20-30 km, such as Mexico (Currie et al., 2002) and Cascadia (Hyndman and Wang, 1995; Gutscher and Peacock, 2003). This depth range could potentially correspond to either a thermal control or to a serpentinized mantle wedge, depending on the geometry and lithology of the margin in question.

For most well studied margins (SW Japan, Cascadia, Chile, Mexico, Alaska), there seems to be a fairly good agreement between the thermally predicted updip and downdip limits, and these limits as obtained from other methods (e.g. - source-time studies, aftershock studies, geodetic studies, tsunami modeling) (Hyndman et al., 1995; Hyndman and Wang, 1995; Oleskevich et al., 1999; Currie et al., 2002; Gutscher and Peacock, 2003) (Figure 3). Thus, thermal modeling appears to be a very useful method for determining the position of both the updip and downdip limit.

## 2 Multi-channel seismic reflection data

During a multichannel seismic (MCS) survey with the German R/V Sonne in January/February 2006 a comprehensive dataset of 5358 line kilometers was acquired coincident with gravity and magnetics in Indonesian waters over the southern part of the 2004 and the 2005 rupture zones. We used a 240 channel, 3 km streamer, and a tuned airgun array consisting of 16 airguns with a total capacity of 50.8 liters. Record length was 14 s with a sample interval of 2 milliseconds.

Processing of the two MCS lines BGR06-115 and BGR06-141 was performed up to full Kirchhoff pre-stack depth migration and included the production and correction via MVA (migration velocity analysis) of a depth velocity model. After testing various combinations of processing parameters the following sequence was regarded as optimal. Pre-stack processing included geometry editing, deconvolution, true amplitude recovery, and time and space variant signal filtering. Reduction of water-bottom multiples was achieved by applying a parabolic radon filter and inner trace mutes. Stacking velocities, at an average distance interval of 3 km, were determined for the reference post-stack time migrated sections. The initial depth model was derived by combining the velocity field from the smoothed stacking velocities (upper section) with the wide-angle/refraction seismic data from the Profile AB located close to these profiles along the lines BGR06-115 & BGR06-141 for the lower section. Particularly the upper parts of the velocity fields were iteratively improved via MVA until the migrated CRP gathers were flat. Quality control included a detailed evaluation of congruence between the post-stack migrated sections and the time-converted pre-stack depth migrated sections.

The MCS data were used to constrain the shallow crustal structure in the trench region and the frontal portion of the accretionary wedge. Profile BGR06-115 images the thick undeformed sedimentary section (3 s TWT) at the trench and the accretionary wedge (Figure 4). The major reflectors observed were used on the one hand to help construct the p-wave velocity model based on the OBS data. For the thermal modeling described below it was also particularly important to obtain the position of the deformation front, the sediment thickness above the decollement and the geometry (depth and dip) of the decollement fault plane, which represents the plate boundary fault.

## 3 Wide-angle seismic data

Wide-angle seismic data were processed using a tomographic approach to obtain an independent velocity model (Zelt and Barton, 1998), followed by forward modeling in order to incorporate secondary arrivals and the main reflectors from the reflection seismic data (Zelt and Smith, 1992). Error analysis included calculation of synthetic seismograms, in order to constrain velocity gradients, and gravity modeling.

### 3.1 OBS data acquisition and quality

During the Sumatra-OBS cruise (R/V Marion Dufresne, July-Aug. 2006) 56 ocean-bottom seismometers from the British OBIC pool, the French INSU pool, the University of Brest and Ifremer were deployed along the above-mentioned SW-NE oriented Sumatra-AB profile just north of Simeulue Island, in the epicentral area (Figure 2). The profile is 252 km-long, resulting in a 4.6 km instrument spacing. All instruments were successfully recovered. A total of 2090 shots were fired on the profile by a 8260 in$^3$ airgun array tuned to single-bubble mode to enhance the low frequencies and allow deep penetration (Avedik et al., 1993). Preprocessing of the OBS data included calculation of the clock-drift corrections to adjust the clock in each instrument to the GPS base time. Instrument locations were corrected for drift from the deployment position during their descent to the seafloor using the direct water wave arrival. The drift of the instruments never exceeded 100 m. Picking of the onset of first and secondary arrivals was performed semi-automatic with a manual quality control and without filtering where possible.

205        Data quality along the profile is generally good on all channels, although some instruments show bands of noise on their geophone probably due to currents. Instruments deployed without a trailing buoy displayed lower noise levels. The OBS's located on oceanic crust show very good data quality (Figure 5) with clear arrivals from the sedimentary layers, crust and upper mantle. Towards the toe of the accretionary prism, the OBS data sections

210    show an increasing amount of sedimentary arrivals (Figure 6). On the accretionary wedge the data sections are highly asymmetric and display long and partly disturbed sedimentary arrivals (Figure 7). Those instruments located in the forearc basin are mostly characterized by clear sedimentary as well as reflections from the base of the top and base of oceanic crust. Data sections from the most landward OBS's on the profile allow us to identify reflected

215    arrivals from the base of the continental crust as well as the base of the oceanic crust that subducts below (Figure 8).

### 3.2 Tomographic inversion

       The tomographic inversion code FAST (Zelt and Barton, 1998) was used to constrain a velocity distribution that was used as an initial guideline to the forward modeling, described

220    hereafter. This code uses a regularized inversion in which user can specify parameters for the travel time misfit and model roughness of the final solution. The method is linearized in that a starting model and iterative convergence scheme are employed. Non-linearity is accounted for by calculating new ray paths for each iteration. The method generates smooth models, which do not resolve sharp boundaries but steeper velocity gradients instead. The most important

225    structural features are thus resolved in an objective manner, i.e., not user-oriented.  Similar tomographic inversion methods have been used in previous studies to constrain the crustal and upper mantle velocities in subduction zones (eg. Graindorge et al., 2003; Dessa et al., 2004; Ramachandran et al, 2006; Gailler et al., 2007) .

       In order to perform this tomographic inversion of the first arrivals, 33 127 travel-times

230    have been picked in the complete data set. Each pick has been assigned a picking error between 20 ms and 150 ms depending on the data quality.  The tomographic model used a grid of 320 km x 40 km with a 1 km grid cell size (Figure 9). For the final model run, 5 different smoothing weights were tested in 10 non-linear iterative steps. The final model predicts a mean travel-time misfit of 156 ms. 97 % of all picks were traced in the model.

### 3.3 Forward ray-tracing modelling

To be able to include reflections from the oceanic and continental crust into the modeling process, we converted the tomographic model into a forward ray-tracing model using the RAYINVR software (Zelt and Smith, 1992) (Figure 10). The number of picks used for the forward modeling was 36322, about 3200 picks higher than the tomographic model. Sedimentary layers were taken from iso-velocity contours of the inversion model and from reflection seismic data of the region of the model. Crustal layers and the Moho were modeled using phases generated by diving waves and deep reflected phases. This was done in accordance with the tomographic model. For the model parametrization, we used the minimum-parameter/minimum structure approach, to avoid inclusion of velocity or structural features into the model unconstrained by the data (Zelt 1999). Velocity gradients and phase identification in the velocity model were further constrained by synthetic seismogram modeling using the finite difference modeling code from the Seismic Unix package (Cohen and Stockwell, 2003; Stockwell, 1999) (Figure 5b, 6b, 7b, 8b).

### 3.4 Error analysis

In order to constrain the dependency of the final tomographic model on the initial model and especially the robustness of the high velocity zone interpreted to be the mantle wedge, which is located at the boundary of the model, different model runs were conducted using different initial models. A variety of simple initial models where selected and the inversion performed (Figure 11). The resulting models are characterized by a lower fit of the data. Nevertheless the high velocity anomaly is found in all resulting models, even though the exact location and amplitude do vary. One test run with high velocities at unrealistically low depths underneath the continental crusts produced a velocity anomaly more shallow (15 km depth) than our preferred final model.

Two-point ray-tracing between source and receiver (Figure 12) shows the well-resolved and the unconstrained areas. Ray coverage for diving and reflected waves is generally very good due to the excellent data quality and close instrument spacing (Figure 12 a and b). All sedimentary layers are well sampled by reflected and turning rays in the model. The crustal layers are well sampled except for the oceanic crust of the subducted slab at depth larger than 30 km. The oceanic Moho at depth greater than 30 km, the continental Moho and 265 the mantle wedge geometry are mainly constrained by reflected arrivals, which generally produce higher amplitude arrivals than diving waves from layers of low seismic velocity gradients such as the upper mantle. The Moho has been additionally constrained by gravity modeling at the ends of the profiles. The fit between predicted arrival times and travel-time picks provides information about the quality of the model (Figure 12). The corresponding

270 misfit is 0.125 s using 95% of the picks.

Construction of a tomographic model as well as a forward ray-tracing model allows us on the one hand to include additional information from the reflected phases and the multi-channel data into the model and on the other hand to verify that all structures from the forward model are required to fit the data. These models are therefore complementary and

275 help to support our conclusions.

As the correct identification of velocity anomalies and the degree of resolution of the model are of fundamental interest to the conclusions of the work, additional calculations of the velocity resolution and depth uncertainty have been performed. The quality of the velocities in the model can be gauged from the resolution parameter (see Figure 13).

280 Resolution is a measure of the number of rays passing through a region of the model constrained by a particular velocity node and is therefore dependent on the node spacing (Zelt, 1999). If a layer can be modeled with one single velocity gradient the resolution parameter will be high even in areas which have lower ray coverage as the area is related to only one velocity node. Nodes with values greater than 0.5 are considered well resolved (Figure 13).

285 In order to estimate the velocity and depth uncertainty of the final velocity model a perturbation analysis was performed. The depths of key interfaces were varied and an F-test was applied to determine if a significant change between models could be detected. The 95% confidence limit gives an estimate of the depth uncertainty of the interface (Figure 13). The sedimentary and crustal layers are well constrained throughout the model. The velocities in

290 the mantle wedge show a resolution of between 0.3-0.7 due to the missing information from turning rays in this layer. The upper mantle velocities are well-constrained. At greater depth the velocities are less constrained due to fewer rays penetrating into this deeper portion of the model. The depth of the main depth interfaces is constrained to a depth error of about +- 1 km.

295        In order to constrain the velocity gradients of the different layers, synthetic seismograms were calculated and compared to the data sections. The finite difference modeling code from the Seismic Unix package (Cohen and Stockwell, 2003; Stockwell, 1999) was used to calculate synthetic seismograms of a record length of 30 s at a 100 m spacing (Figure 5b, 6b, 7b, 8b). The program uses the explicit second order differencing method for

300       modeling the acoustic wave equation. The input velocity model was calculated from sampling the forward velocity model at a lateral 50 m interval and 10 m interval in depth. In order to avoid grid dispersion, the peak frequency of the Ricker wavelet source signal is calculated to be equal to the lowest velocity of the medium divided by the grid points per wavelength multiplied by 10. In this case the source wavelet is centered at 8 Hz, similar to the signal from

305       the airgun array used during the cruise. The boundary conditions were set to be absorbing at the sides and bottom of the model and free at the surface. Detailed ray-coverage, travel-time fit and synthetic data provide additional information of the identification of the picked phases (Figure 14 and 15).

### 3.5 Gravity modeling

310       Since seismic velocities and densities are well-correlated, gravity modeling provides an important additional constraint on the seismic model. Areas unconstrained by the seismic data can be modeled by comparing calculated gravity anomalies with those observed.

        The gravity data were forward modeled using the gravity module of the software of Zelt and Smith (1992). Average P-wave velocities for each layer of the seismic models were

315       converted to densities in good agreement with different velocity density relationships from laboratory measurements (Ludwig, Nafe and Drake (1970), Carlson and Herrick, 1990, Hughes et al., 1998 and Hamilton et al., 1978) for the sedimentary layers and the relationship of Christensen and Mooney (1995) for crustal layers (Figure 16). Layers from the velocity modeling have only been subdivided where necessary and strong lateral velocity gradients are

320       present. The upper mantle densities were set to a constant 3.32 g/cm$^3$. To optimize the fit of the final gravity model (Figure 17) the densities of each layer were subsequently manually varied within an error bound of 0.25 km/s from their original value. This deviation is considered a realistic uncertainty of empirical relationships used for the velocity-density conversion.  To avoid edge effects both models were extended by 100 km at both ends and 325  down to a depth of 95 km. The calculated anomalies can be compared with the shipboard measured gravity anomaly (Figure 17). The predicted anomalies generally fit the observed data well. The largest misfit is observed at around 240 km model distance and might be caused by three-dimensional effects of the basement topography.

For comparison we have calculated two additional models, in which the mantle wedge
330  is replaced by either continental crustal material, with a density of 2.83 kg/m$^3$ or normal upper mantle material characterized by a density of 3.32 kg/m$^3$ (Figure 17). Neither of the two alternative models shows a fit as satisfactory as our preferred model with a density of 3.1 kg/m$^3$

## 4 Numerical modeling of the Sumatra fore-arc thermal structure

335  We applied finite-element modeling of forearc thermal structure in order to determine the temperature distribution along the plate interface and to predict the updip and downdip limits of the seismogenic zone. This approach is based on models, which consider temperature (together with lithology) as one of the primary controls of stick-slip rheological behavior that lead to earthquake rupture (Hyndman and Wang, 1993; Hyndman et al., 1997; Peacock and
340  Wang, 1999; Gutscher and Peacock, 2003) (Figure 2). The geometry of the subducting oceanic crust and the upper plate (down to 20-30 km depth) is constrained primarily by the OBS data presented above. The total sedimentary thickness of 5 km at the trench and a decollement at 4 km depth are taken from published seismic reflection data (Karig et al., 1980; Singh et al., 2008) and the BGR seismic data presented here (Franke, et al, 2008) (Fig.
345  4). The deeper geometry is obtained from the distribution of Wadati-Benioff zone hypocenters. The global relocated hypocenter catalog (Jan. 1964 - Dec. 1995) was used together with 3 additional years of data available online (Engdahl et al., 1998). Additionally data from relocated aftershocks were taken to help constrain the plate interface in the 30-50 km depth range (Engdahl et al., 2007). Together, these data are used to construct the 2-D
350  finite element grid.

We used finite element (FE) software developed by Kelin Wang (Wang et al., 1995). The 500 km long FE-models consists of 936 quadrilateral elements, with a total of 2933 nodes. The models include the effects of radiogenic heating in the crust, shear heating along the subduction interface (to a distance of 240 km from the trench) for an effective shear stress

355   of 10 MPa, and viscous corner flow in the mantle wedge. Thermal conductivity in the mantle
      and oceanic crust is 3.138 W/mK consistent with the GDH1 model (Stein and Stein, 1992)
      and thermal conductivity in the continental crust is 2.5 W/mK (Peacock and Wang, 1999).
      Heat generation in the upper continental crust is 2.5 microWatt/m2 and in the lower
      continental crust is 0.27 microWatt/m2. Models with lower radioactive heat generation (1.3

360   microWatt/m2) resulted in nearly the same thermal structure at depth (along the plate
      interface), but predict a significantly lower surface heatflow than that observed in the arc and
      backarc region. Thus, the higher value was selected in agreement with recent work (Hippchen
      and Hyndman, 2008). The model geometry is shown in Figure 18.

      The initial boundary conditions include: at the left side, oceanic lithosphere isotherms

365   for a subducting oceanic plate of the appropriate age based on the GDH1 thermal cooling
      model (Stein and Stein, 1992), 0° C at the surface and an appropriate continental geotherm at
      the right side boundary (representing the upper plate). The three primary input parameters to
      the model are thus; the plate geometry, the age of the subducting lithosphere and the
      subduction velocity.

370   The modeled transect is located on the NW Sumatra margin, in the epicentral region of
      the 26 Dec. 2004 earthquake. The age of the subducting oceanic lithosphere is known from
      magnetic anomaly studies (Mueller et al., 1997) and is approximately 60 Ma. At the northern
      tip of Sumatra, the relative plate motion between the Australian plate and the Sunda block is 5
      cm/yr in a N8°E azimuth and the motion between the Indian plate and the Sunda block is

375   4cm/yr in a N20°E direction (Vigny et al., 2005). The component of plate motion
      perpendicular to the margin thus yields an orthogonal subduction velocity of 3cm/yr, which
      was used as the preferred velocity for the thermal modeling.

## 4.1 Comparison of observed and calculated heat flow

380   Heat flow data are available in the study area from several different sources. Older
      marine and terrestrial heat flow values were obtained from the Global Heat Flow Database (
      http://www.heatflow.und.edu ). Marine heat flow data were acquired during the Marion
      Dufresne Aftershocks cruise (Sibuet, 2005) and are shown here. Two other heat flow studies
      performed during the R/V Sonne cruise 189, one with in-situ measurements, and one obtained

385    from BSR observations (calculated using gas-hydrate stability conditions) are also included here (Delisle and Zeibig, 2007). The observed heat flow pattern (Figure 19, A) shows a fairly high degree of scatter, but some general trends are discernable. In the undisturbed oceanic domain, heat flow is about 60 mW/m2. Three lower heat flow values near the trench (30-40 mW/m2) are likely due to the cooling effect of hydrothermal fluid circulation as observed in

390    other subduction zone environments (Grevemeyer et al., 2005, EPSL, 236, 238-248). Here the effect appears to be local and not representative of large-scale lithospheric cooling as described for the Central American trench (Harris and Wang, 2002), since the mean heat flow in this region (from the oceanic crust to the toe of the wedge) is about 60 mW/m2, the value expected for typical 60 Ma oceanic lithosphere. Heat flow in the fore arc generally declines to

395    values in the 40-60 mW/m2, though there is wide scatter in the observed data, with some measurements as high as 80 mW/m2. An increase in heat flow is observed in the arc and back arc region to values of 80-90 mW/m2, as reported for other back arcs (Currie and Hyndman, 2006).

       The expected heat flow at the surface calculated using the thermal model presented

400    here is shown in Figure 19, A. The calculated 60mW/m2 at the trench is in fairly good agreement with the observed data. A decline to about 40mW/m2 is predicted in the fore-arc, followed by an increase to 80mW/m2 in the arc region. No data are available to check the modeled heat flow predicted for the on-land portion of the forearc.

405    **4.2 Thermal structure and the seismogenic zone**

       The modeled thermal structure is presented in Figure 19 B. The OBS velocity model as well as the hypocenters used to constrain the modeled geometry, are shown as well. On the basis of the 100-150°C isotherms, the updip limit is predicted to be very close to the trench (within 5-30 km). This appears to be due to the insulating effect of the thick sedimentary

410    cover at the trench. The thermally predicted position of the updip limit is in good agreement with the observation of numerous aftershocks with a shallow thrusting mechanism in this zone (Fig. 14C) (Engdahl et al., 2007). These results are also in agreement with a recently published thermal model of the NW Sumatra margin located roughly 100-200 km further SE (Hippchen and Hyndman, 2008) which also predicts an updip limit near the trench. Their

415    150°C isotherm is 34 km from the trench. For the range of models we tested, we obtained a

150°C isotherm 30±10 km from the trench. Our 350°C and 450°C isotherms, which are considered to correspond to the downdip limit, are located 210 km and 250 km, respectively,

from the trench axis. The thermal model further SW (at the limit between the 2004 and 2005

earthquakes) predicts the location of the 350°C and 450°C isotherms at distances of 214 km

420    and 254 km, respectively (Hippchen and Hyndman, 2008), which is nearly identical to the results we obtain. Aftershocks are observed near the shallow dipping plate boundary up to a distance of 220 km from the trench (Engdahl et al., 2007), which is once again in good agreement with the thermally predicted limits.

## 5 Discussion

425          Tomographic inversion and forward modeling of wide-angle seismic data from a line

located close to the epicenter zone of the great earthquake of December 2004 allows imaging

of the accretionary prism and subducted slab to a depth of 35 km (Figure 9 and 10). The

accretionary prism shows sedimentary layers with a total thickness of 20 km. 4 km of

sediments overlie the oceanic crust at the trench, probably consisting of hemi-pelagic

430    sediments and trench fill. The seismic velocities of the deeper layers are up to 5.5 km/s high,

probably due to compaction and to some low to middle-grade metamorphic reactions.

Sedimentary velocities of the forward model are taken from iso-velocity contours of the

tomographic model. They roughly correspond to lines of equal lithostatic pressure.  As one

single sedimentary layer recognized in the MCS data might be characterized by lower

435    velocities where its position is shallow and higher velocities in regions where its position is deeper, the velocity contours do not coincide exactly with the deep sedimentary layer stratification imaged by the reflection seismic data. In the Simeulue fore-arc basin the iso-velocity contours show a pronounced depression, following in general the Neogene basin fill geometry as described by Berglar et al (2008).

440          The oceanic crust imaged by the reflection or wide-angle seismic data is only about 5

km thick (Franke et al. 2008; Singh et al., 2008; this paper) and substantially thinner than

"normal" oceanic crust, which is characterized by a thickness of 7.1 +- 0.8 km (White et al.,

1992). Velocities range from 5.80 to 6.80 km/s at the trench to 6.20 - 6.60 km/s at 270 km model distance. The oceanic Moho found from wide-angle seismic modeling is located at the 445    lower boundary of a series of prominent, discontinous reflectors in the reflection seismic data. Unusually thin oceanic crust is known to form at ultra slow spreading centers, where serpentinized mantle material commonly outcrops at the sea floor (Jackson et al., 1982; Muller et al., 1997; Jokat and Schmidt-Aursch, 2007). The oceanic crust currently subducting beneath the NW Sumatra margin was formed around 50-60 Ma at the Wharton spreading

450    center at a half spreading rate between 50 and 75 mm/a (Royer and Sandwell, 1989). Thus, very slow spreading can be ruled out to have formed the thin oceanic crust.  Alternatively the thickness of the crust might be explained by an unusually low mantle potential temperature at the spreading center leading to low degrees of partial melting in the mantle (Klein and Langmuir, 1987).

455    The velocities in the oceanic upper mantle are reduced at model distances between 0 and 140 km, with values of 7.40 to 7.60 km/s (Figure 10). A similar reduction in upper mantle velocities was observed offshore Costa Rica over the flexurally faulted portion of the oceanic Cocos plate before it enters the Middle America trench and interpreted as being due to serpentinization of the uppermost mantle (Grevemeyer et al., 2007) through faults generated

460    by the flexure and imaged on seismic reflection data (Ranero et al., 2003). The oceanic crust at this location offshore Central America formed at the fast spreading East Pacific Rise and is unusually thin (4.8 – 5.5 km thick) (Grevemeyer et al., 2007). These thicknesses are very close to those obtained by modeling of the OBS profile off NW Sumatra presented here (Fig. 10) and are also observed by the MCS data (Fig. 4). It is likely that thin oceanic crust may be

465    fractured more readily and therefore may permit a higher degree of serpentinization in the upper mantle than thick oceanic crust.

We find a strong reflection from a crustal unit at a distance starting 120 km (220 km model distance) from the trench and extending landwards (Figure 15). This backstop is characterized by a rough surface showing several blocks and velocities between 6.30 km/s at

470    its top and 6.80 km/s at its base. These velocities and the relatively low velocity gradient of this layer can be characteristic of either igneous continental crust (Christensen and Mooney , 1995) or highly metamorphosed sediments. The corresponding continental Moho is located at a shallow depth of 22 km, similar to that proposed by Kiekhefer et al., 1980 from wide-angle seismic modeling of trench parallel profiles near Nias island and by Simoes et al., 2004 from 475  gravity modeling of a profile south of Nias island. Both of these studies are located in the segment ruptured during the March 2005 earthquake.

An earlier study based solely on multi-channel seismic data acquired by Western Geco, draws the conclusion that the 26 Dec. 2004 earthquake ruptured along a shallow dipping fault beneath the oceanic Moho (Singh et al., 2008). Both the Western Geco seismic
480  line and the seismic data presented in Figure 4 show clear evidence of folding and thrusting of recently deposited trench fill sediments at the toe of the accretionary wedge. Several of these thrust faults, showing both landward and seaward vergence, can be traced down to a decollement level within the sediments, roughly 1 km above the oceanic basement. These observations suggest that the vast majority of strain from relative plate motion is
485  accommodated here, along the decollement beneath the accretionary wedge and above the downgoing oceanic crust, as observed in all subduction zones worldwide. Mega-thrust earthquake rupture along this decollement and deep into the sub-fore arc mantle offers the most coherent explanation for the 26 Dec. 2004 event.

The thermal model presented here predicts a seismogenic zone extending down to 40
490  km depth and a distance of 220 km from the trench. On the basis of the OBS velocity model, this is well below the forearc crust (which has a thickness of about 20 km at a distance of 170 km from the trench). This structural peculiarity is ascertained by several pieces of evidence and supporting elements: (i) the mantle velocity anomaly below the forearc that is shown to be robust through several tomographic inversion runs starting from fairly different initial
495  models and also (ii) to be necessary to model reflections from the oceanic plate subducting below (Fig. 14 and 15); (iii) the observation of reflected phases corresponding to this shallow Moho that can be modeled as well (Dessa et al. 2009); (iv) the report of a similarly shallow mantle anomaly on favorably oriented profiles in the Nias basin, some 350 km south of our study zone (Kieckhefer et al. 1980).

500  Together with the seismic data on deep structure, the thermal modeling permits us to assess the hypothesis that either temperatures of 350-450°C or the sub-forearc Moho defines the downdip limit of the seismogenic zone, whichever is shallower (Hyndman et al., 1997). The seismic velocity model clearly indicates that the Moho is shallower than the observed downdip extent of the seismogenic zone (as defined by aftershocks). This does not support the

505     hypothesis that the Moho is its downdip limit. Our results imply that a significant portion
        (>50 km) of the rupture occurred along the interface between the oceanic crust of the
        downgoing plate and the fore-arc mantle of the upper plate. This contradicts the conclusions
        of a recent thermal modeling study, which suggested that the 30 km deep downdip limit was
        controlled by the presence of a serpentinized mantle wedge beneath the continental Moho
510     (Hippchen and Hyndman, 2008). Indeed until now, the 1994 Sanriku-Oki earthquake was the
        best documented example of a subduction earthquake which ruptured into the sub fore arc
        upper mantle (Hino et al., 2000). Here, in the Northeast Japan subduction zone, a M7.7 event
        occurred and the detailed aftershock distribution was obtained by deploying a network of
        OBS on the seafloor. The aftershocks extended to 50 km depth and the upper plate Moho is
515     known from deep crustal seismic studies to be located at 20 km depth (Hino et al., 2000). A
        recent review of the seismogenic zone around the Japanese Islands confirms this tendency
        (deep, sub-Moho rupture) for the northern Honshu and Kanto portions of the Japan trench
        (Seno, 2005). The recent Tokachi-oki M8 earthquake of 2003 offshore Hokaido also exhibited
        very deep rupture (40-55 km) (Machida et al., 2009).

520         Our thermal model also predicts a very shallow updip limit of the seismogenic zone
        extending almost all the way to the trench and thus implies a very large downdip width of the
        seismogenic zone (about 200 km). While the exact extent of the updip portion of the rupture
        plane is difficult to determine precisely (Wang and He, 2008), the results obtained using
        several different methods all agree that the southernmost portion of the 2004 rupture zone, off
525     NW Sumatra was the widest, about 200 km. These methods include source time studies of
        fault slip along the plate boundary fault (Ammon et al., 2005), fault slip inversions using
        geodetic data (Vigny et al., 2005) and tsunami inversions (Fujii and Satake, 2007). The great
        downdip width of 200 km is in agreement with that predicted by our thermal modeling and
        can partly explain the great contribution to seismic moment, due to the large surface area of
530     the fault plane.

## 6 Conclusions

Combined modeling of wide-angle seismic and reflection seismic data of a profile located
close to the epicentral area of the great Sumatra earthquake reveals the crustal structure of the
subduction zone down to a depth of 35 km. The main structures imaged by this data set are:

1) 4-5 km of sediments overlying the oceanic crust at the trench, probably consisting of
older sediments and trench infill.

2) An anomalously thin oceanic crust about 5 km thick. Although most thin oceanic crust
forms at very slow spreading centers, the oceanic crust in this region formed around
50-60 Ma ago at the Wharton spreading center at a fast spreading rate.

3) A backstop structure located about 120 km from the trench beneath the forearc basin
and characterized by seismic velocities and gradients characteristic of continental
crust.

4) A shallow continental Moho at only around 22 km depth, 170 km from the trench.

5) The hypocenter of the great 2004 earthquake is located at the interface between the
downgoing plate and the upper plate continental mantle, indicating that the mantle
wedge is not serpentinised to a degree sufficient to prevent earthquake nucleation.

Thermal modeling of the subduction zone was performed using the crustal structure from
the wide angle seismic data and the distribution of aftershocks. These results indicate:

1) The upper limit of the seismogenic zone (as defined by the 100-150°C isotherms) is
located close to the trench (within 5-30 km). This is in good agreement with the
observation of numerous aftershocks with a shallow thrusting mechanism in this zone.

2) The 350°C and 450°C isotherms are located 210 km and 250 km, respectively, from
the trench axis. This corresponds to the landward limit of aftershocks along the fault
plane. This limit is about 50 km further landward and 18 km deeper than the sub-
forearc Moho.

3) A significant portion (>50 km) of the rupture occurred along the interface between the
oceanic crust of the downgoing plate and the fore-arc mantle of the upper plate. Thus,
the downdip limit of seismogenic rupture off NW Sumatra is not controlled by the
sub-forearc Moho, but appears to be controlled by temperatures of 350-450°C
isotherms, at which felsic rocks begin to deform by ductile flow.

**Acknowledgments**

Oceanographic work in Indonesia was made possible thanks to the strong implication of LIPI (Indonesian Institute of Sciences) and BPPT (Agency For the Assessment and Application Technology). French ANR (Agence National de Recherche) and INSU (Institut National des Sciences de l'Univers) and Ifremer (Institut français de recherche pour l'exploitation de la mer) contributed to the funding of the SUMATRA-OBS cruise. We would like to thank the IPEV (Institut Paul-Emile Victor), for supporting this work. We thank the chief operator H. Leau, the captain F. Duchênes and the crew of the R/V Marion Dufresne for their professional work during the oceanographic expedition. We would also like to thank the technical teams of the four OBS pools (INSU, Ifremer, University of Brest and OBIC) and the Genavir technical team for their professional work during the deployment of ocean bottom instruments and the airgun array. We also thank the master and crew operating RV SONNE during MCS data acquisition. The German Ministry for Research and Education (BMBF) supported the study (grants 03G0186A). The GMT (Wessel and Smith, 1995) and Seismic Unix software package (Stockwell and Cohen, 1999) were used in the preparation of this paper. We thank Bob Engdahl for providing his relocated teleseismic events in the Sumatra region.

We thank Roy Hyndman and an anonymous reviewer and the associate editor Kelin Wang for their exhaustive and highly detailed reviews and for the opportunity to improve the manuscript accordingly.

**Figure captions**

Figure 1: General location map of Sumatra convergent margin. The rupture planes of the three great earthquakes are shown, with the 3-month aftershocks in each case; M9.3 26 Dec. 2004 (red/circles), M8.7 28 March 2005 (yellow/squares), M8.4 12 Sept. 2007 (green/triangles). The stars show where ruptures initiated. Estimated rupture zones from previous great earthquake sequence are shown as dashed lines (Lay et al., 2005). The location of the geophysical transect investigated in this study is indicated by the black line.

Figure 2: Seafloor bathymetry of the study region from Aftershocks and Sumatra-OBS cruises. OBS are marked by red dots and MCS profiles by red lines. Shaded areas are not covered by shipboard measured bathymetry and show predicted bathymetry from satellite

595     altimetry (Sandwell and Smith, 1995). Star marks location of the great Sumatra earthquake.

Figure 3: Schematic representation of a subduction zone showing the seismogenic zone where stick-slip behavior is believed to occur between the 150°C and 350°C isotherms. Note transition zones exist (dashed) above the up-dip limit to around 100°C and below the downdip limit to around 450°C (see text for further explanation).

600     Figure 4: Multichannel reflection seismic profiles from the BGR SEACAUSE II cruise in 2006. (A) Pre-stack depth migrated section of Profile BGR06-115 (B) Pre-stack depth migrated section of Profile BGR06-141 (C) Profiles BGR06-115 and BGR-141 with layer boundaries and velocities from the wide-angle seismic modeling overlain.

Figure 5: (a) Bandpass filtered (3-5 Hz, 24-36 Hz) vertical geophone data section from

605     OBS 05 located on oceanic crust. The data are displayed with a gain proportional to source-receiver offset and are reduced at a velocity of 8 km/s. PmP (reflection from the Moho), and Pn (turning waves from the upper mantle) are annotated (b) Synthetic seismograms calculated from the velocity model for the same station using the finite-difference modelling code from the Seismic Unix package (Cohen, 2003; Stockwell, 1999). The synthetic seismograms are

610     calculated every 100 m with a source frequency centered around 5 Hz.

Figure 6: (a) Data from the vertical geophone data section from OBS 11 located on oceanic crust close to the accretionary prism. The same gain, filter and scaling have been applied as in Figure 4a. (b) Corresponding synthetic seismograms calculated from the model using the same method as in Figure 4b.

615     Figure 7: (a) Data from the vertical geophone data section from OBS 34 located on the accretionary prism. The same gain, filter and scaling have been applied as in Figure 4a. (b) Corresponding synthetic seismograms calculated from the model using the same method as in Figure 4b.

Figure 8: (a) Data from the vertical geophone data section from OBS 44 located the end of the

620     profile. The same gain, filter and scaling have been applied as in Figure 4a. (b) Corresponding synthetic seismograms calculated from the model using the same method as in Figure 4b.

Figure 9: Result of the tomographic inversion of first arrivals. Overlain black lines represent layer boundaries from forward modeling. Circles mark aftershock locations (Engdahl, 2007). The hypocenter of the 2004 earthquake is projected onto the model and marked by a cross.

625     Figure 10: Final velocity models for the Profile including the model boundaries used during inversion (solid lines) and iso-velocity contours every 0.25 km/s. OBS locations are indicated by red circles. Areas unconstrained by raytracing modelling are shaded. Vertical ex. 4.

Figure 11: Variation of the starting model. (A) initial model as used in this study (B) initial model with a 5 km lower crust-mantle boundary (C) initial model with a 5 km higher crust-mantle boundary (D) initial model with strong slope dipping towards the continent in the crust-mantle boundary (E) initial model with strong slope dipping towards the oceanic plate in the crust-mantle boundary (F) resulting velocity model from (A) leading to a ch$^2$ = 9.8 (G) resulting velocity model from (B) leading to a ch$^2$ = 20.25 (H) resulting velocity model from (C) leading to a ch$^2$ = 11.5 (I) resulting velocity model from (D) leading to a ch$^2$ = 14.7 (J) resulting velocity model from (E) leading to a ch$^2$ = 12.6.

Figure 1:Figure 12: (A) Upper panel: Ray coverage of diving waves with every fourtieth ray from two-point ray-tracing plotted. Lower panel: Observed traveltime picks and calculated travel times (line) for the same phases for all receivers along the model. (B) Same as (a) but for reflected phases.

Figure 13: Resolution parameter for depth nodes of the velocity model. Contour interval is 0.1. The depth uncertainty of the most important boundaries calculated from the 95 \% confidence limit of the f-test given in the framed boxes.

Figure 14: (a) Ray-coverage of OBS 26 in the model (b) travel time fit of the associated rays (c) data section corresponding to the model (d) synthetic seismograms corresponding to he model.

Figure 15: (a) Ray-coverage of OBS 28 in the model (b) travel time fit of the associated rays (c) data section corresponding to the model (d) synthetic seismograms corresponding to he model.

Figure 16: Relationship between velocity and density from various publications. Inverted triangles mark velocity and corresponding densities used for gravity modeling. Gray shaded areas mark the error bounds of 0.25 km/s.

Figure 17: Results from gravity modeling. (A) Gravity model with densities used for modelling in g/cm$^3$ indicated by italic numbers. Positions of OBSs (circles) are indicated. Black lines represent layer boundaries from seismic modelling. (B) Shipboard measured free-air gravity anomaly (black line). Predicted anomaly (dashed line) for our preferred model marked by a dotted line, a model in which the mantle wedge has been replaced by continental

crust (2.83 kg/m3 marked by a broken line) and a model where the mantle wedge has been replaced by normal mantle material (3.32 kg/m3, marked by solid line).

Figure 18: Finite element grid used for calculating forearc thermal structure. The three most important factors controlling the thermal structure are the age of the subducting oceanic lithosphere, the convergence velocity, and the geometry of the subduction zone. The model includes radiometric heating in the crust, shear heating along the plate boundary and convection in the asthenospheric wedge beneath the arc (using a Bachelor's corner flow solution).

Figure 19: Thermal model of the Sumatra subduction zone. (A) Location of the thermal modeling profile (red line), earthquake hypocenters used to construct the model geometry (colored circles), heat flux measurements (colored diamonds) and active volcanoes (triangles) (B) Heat flux measurement (diamonds) and calculated heat flux from depth of a bottom simulating reflector (BSR) (light blue zone) and calculated heat flow (black line) shown along transect from thermal model shown in B. (C) Thermal structure along the transect (shown in Fig. 1). The geometry is constrained by the reflection and wide-angle seismic data in the shallow portion (0-20 km depth) and by earthquake hypocenters (circles) in the deeper portion. The thermally predicted seismogenic zone has a downdip width of roughly 200 km (180 km taking only the 150°C and 350°C isotherms, 240 km when including the transition zones to 100°C and 450°C). The gray shaded regions indicate the variation in the horizontal (downdip) position of the 150°C, 350°C and 450°C isotherms for the range of models tested (v = 2, 3 and 4 cm/yr and slab ages of 50, 60 and 80 Ma) and is thus a measure of uncertainty. Note the 100°C isotherm shifts less than 2 km for all models tested. The updip limit extends very close to the trench, beneath most of the accretionary wedge. The downdip portion extends well into the fore-arc mantle of the upper plate. The thermally predicted limits are in good agreement with the observed distribution of relocated aftershocks (black squares) (Engdahl et al., 2007).

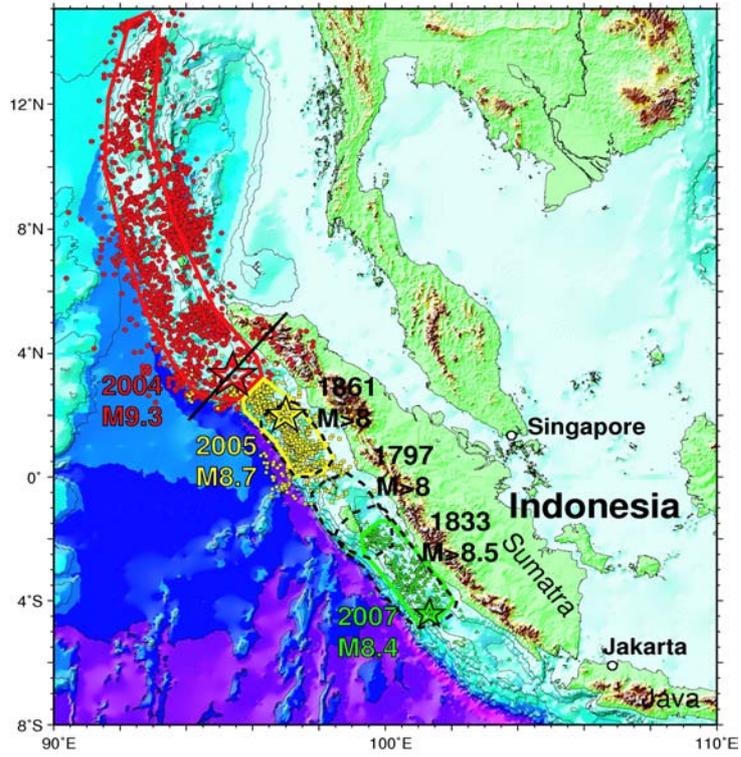

945    Figure 1

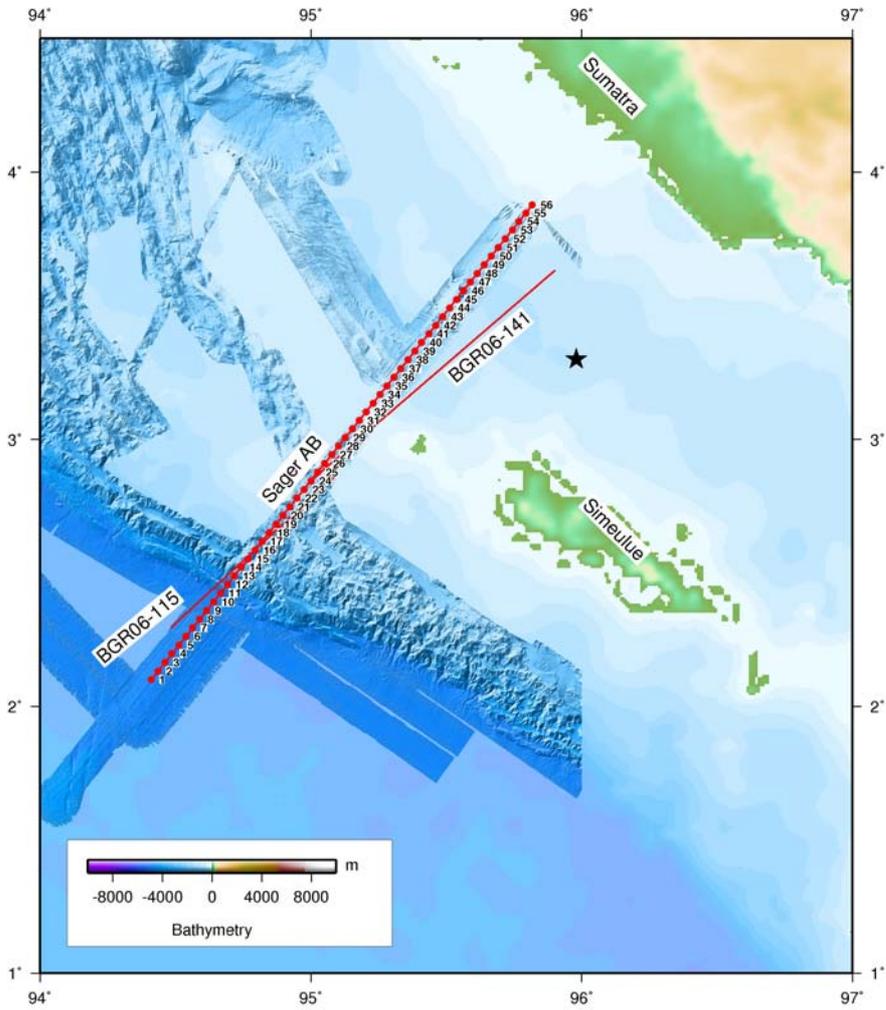

Figure 2

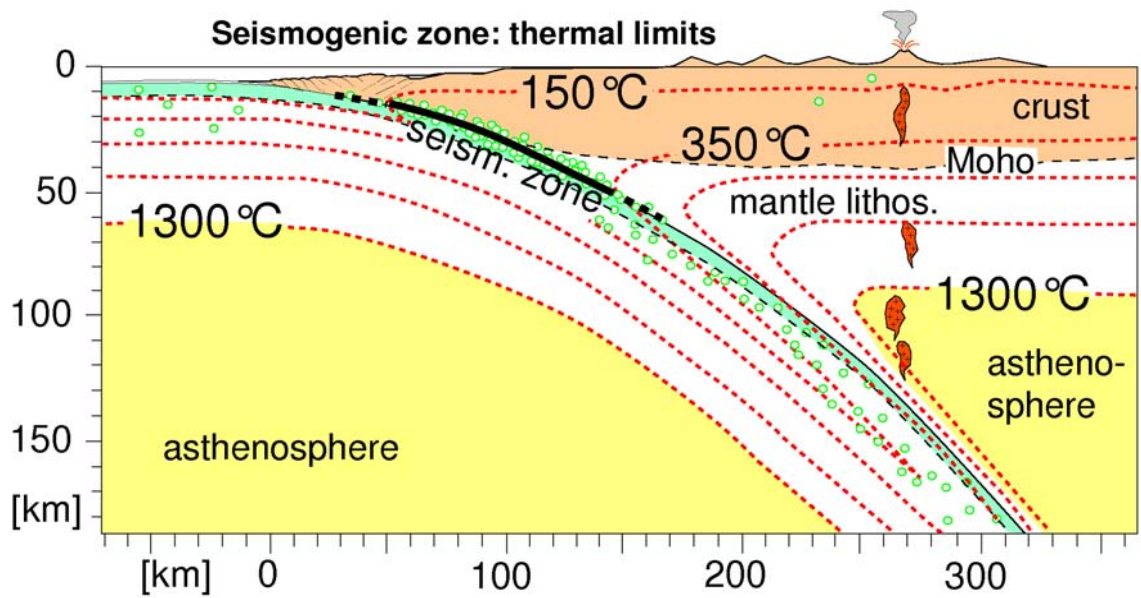

    Figure 3

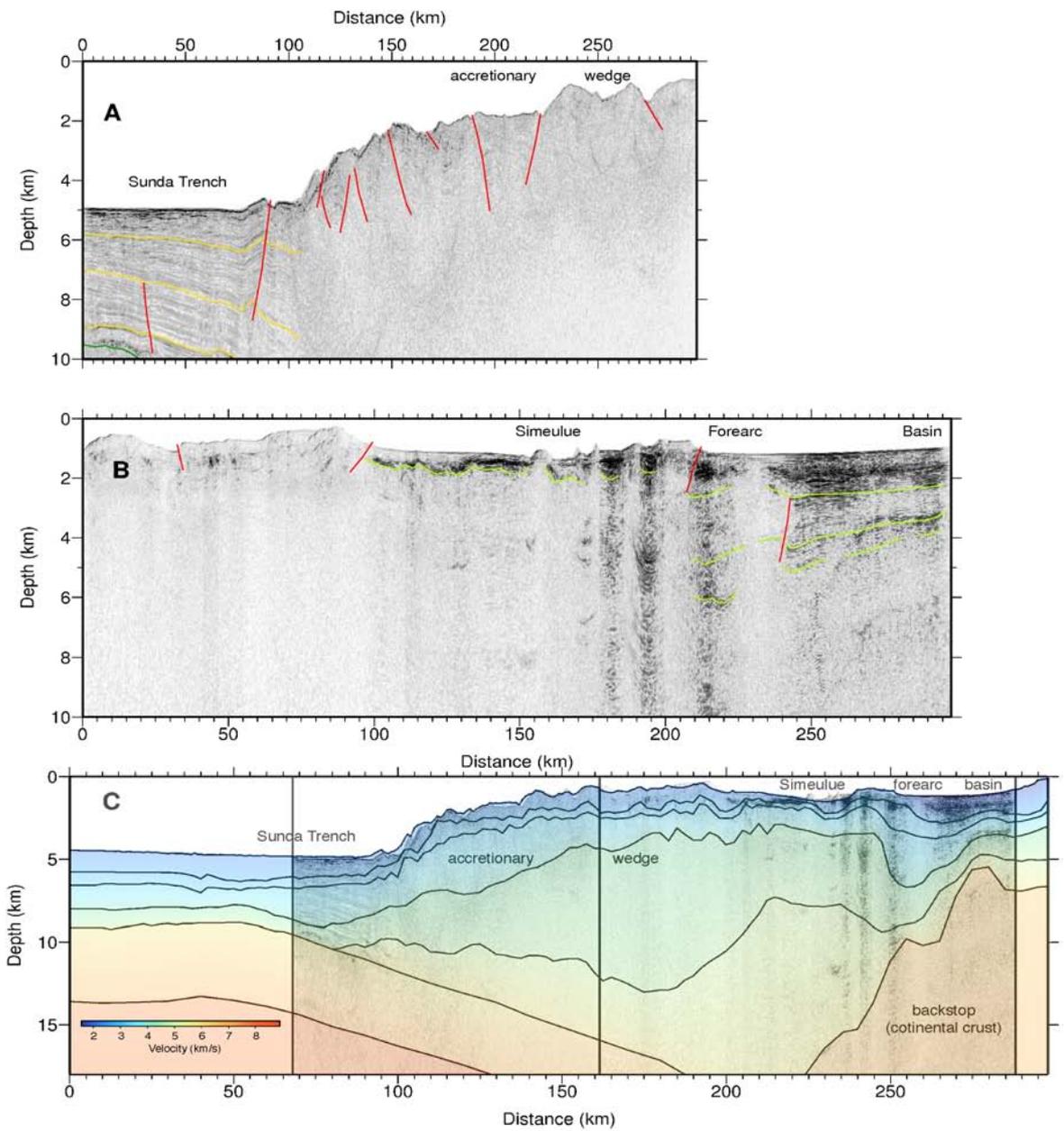

Figure 4

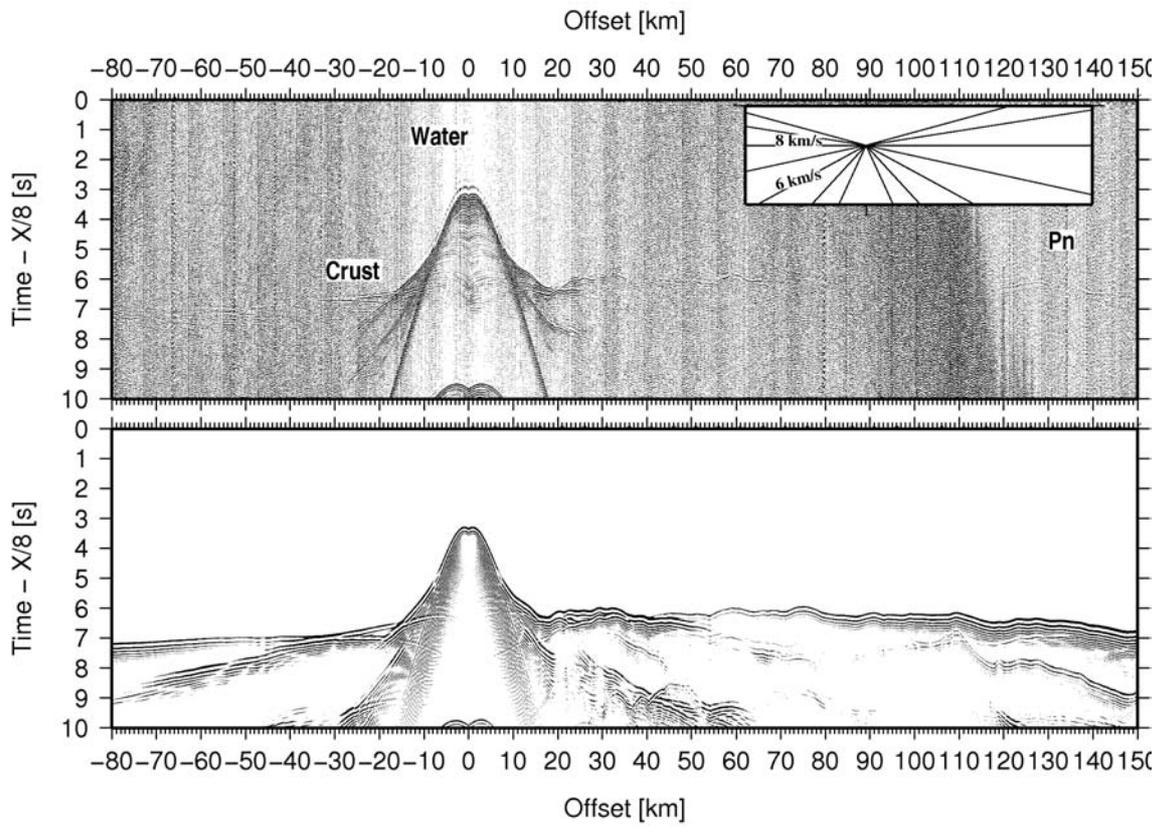

Figure 5

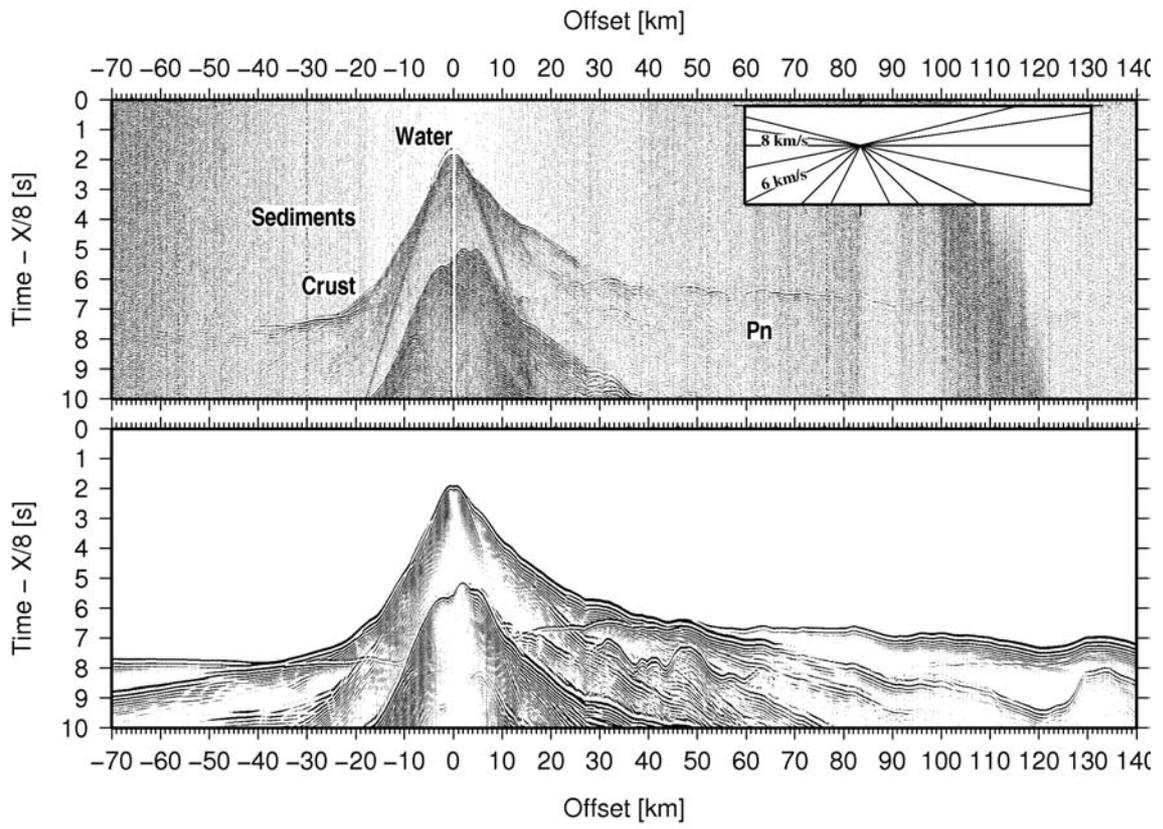

955

Figure 6

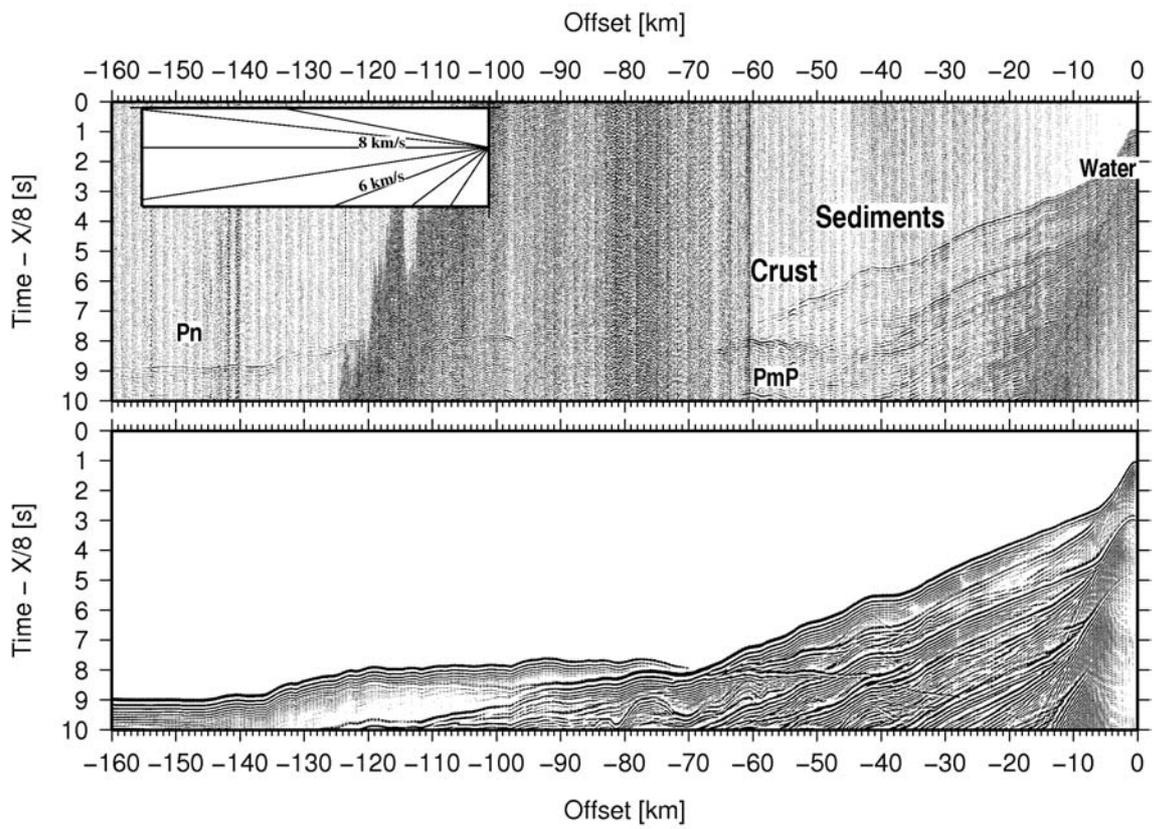

Figure 7

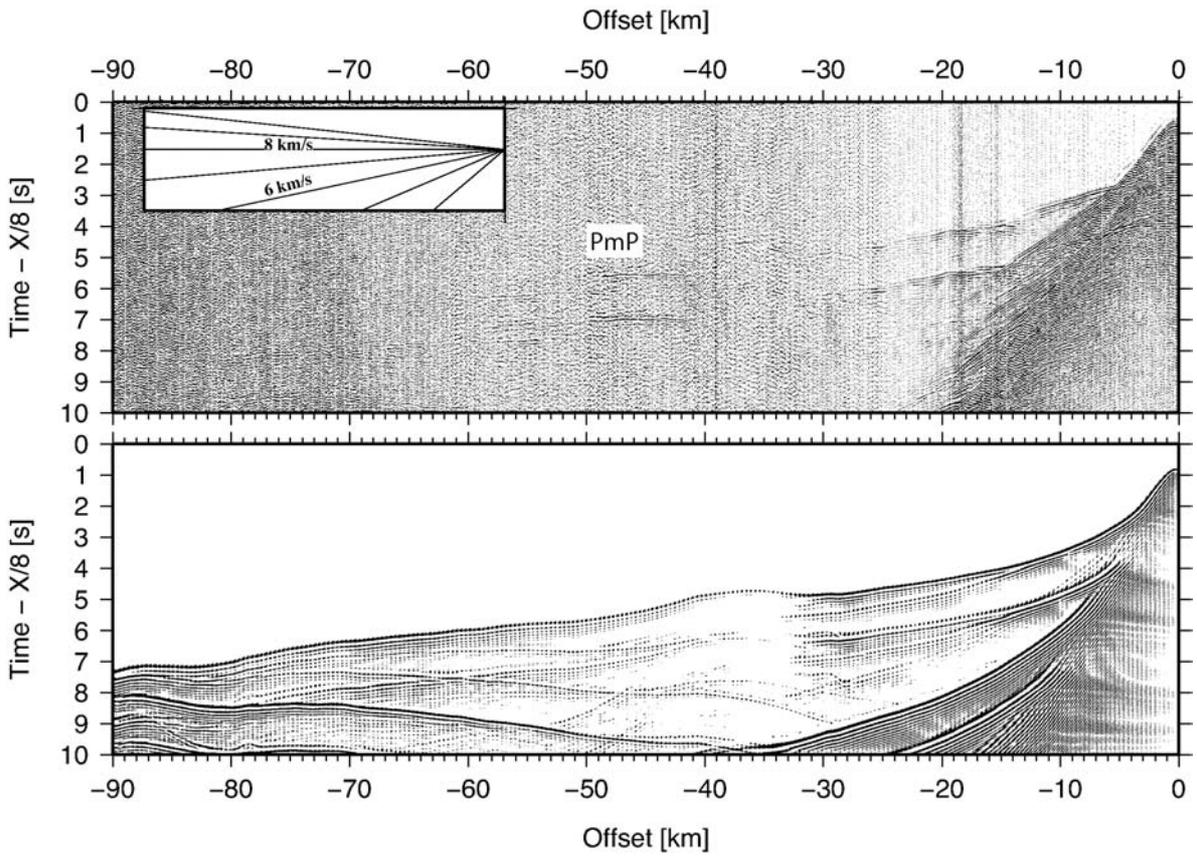

960    Figure 8

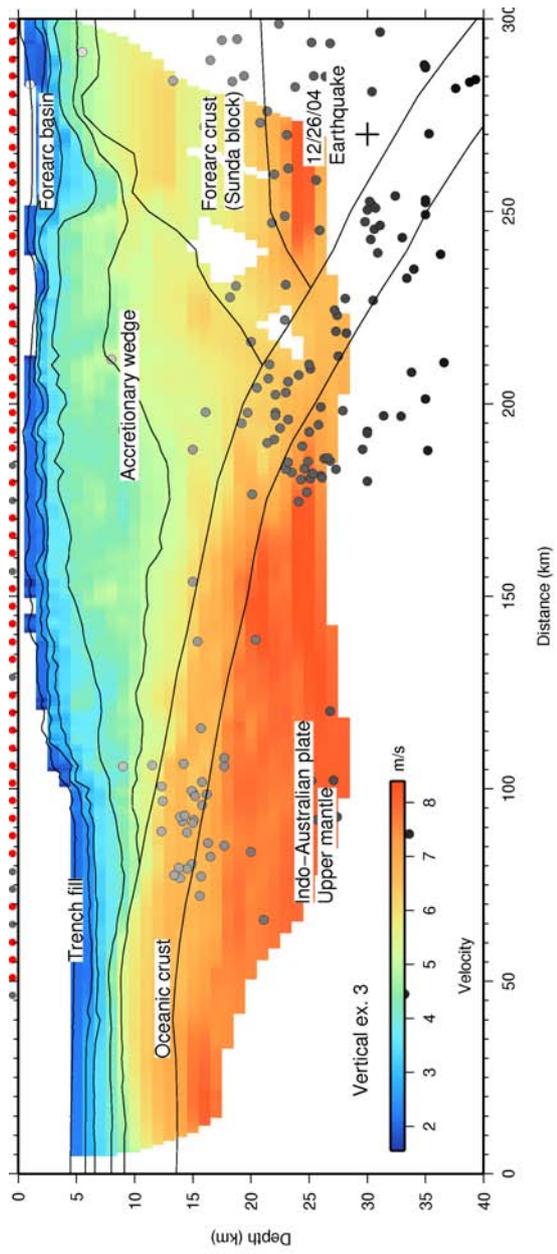

Figure 9

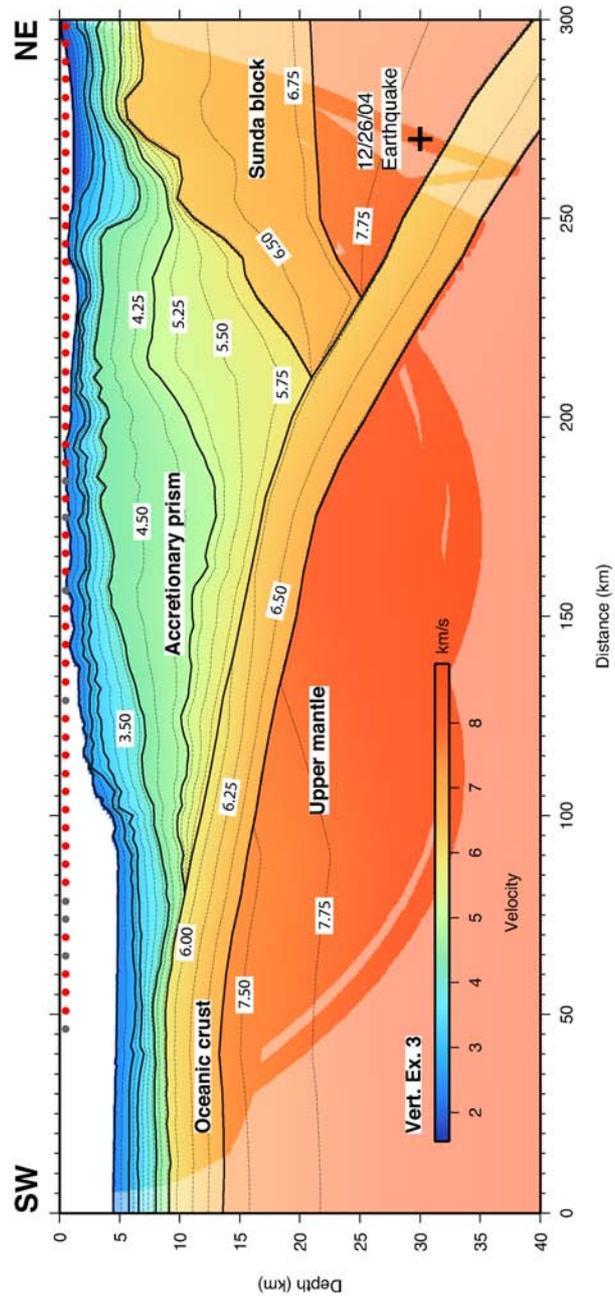

Figure 10

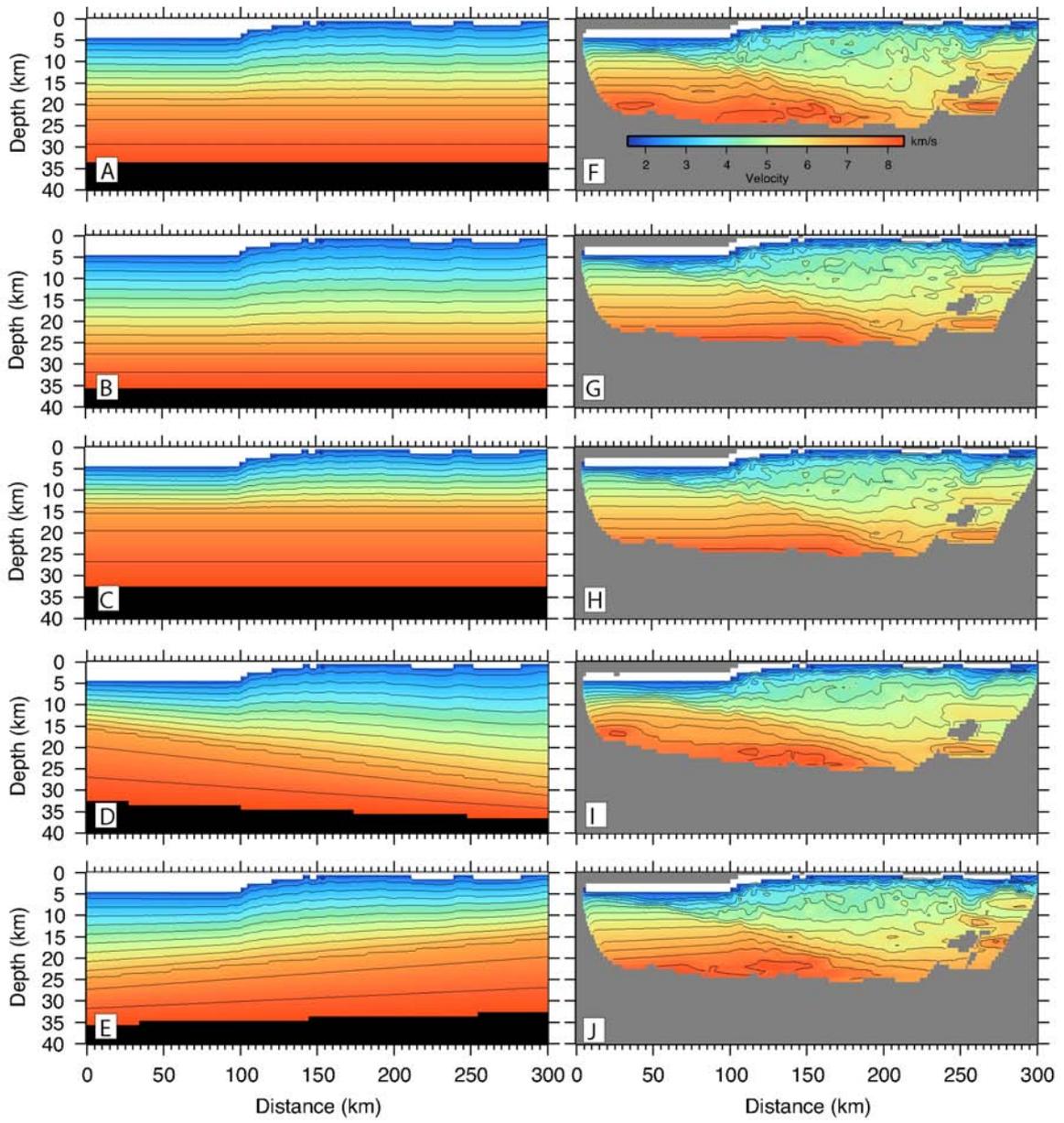

Figure 11

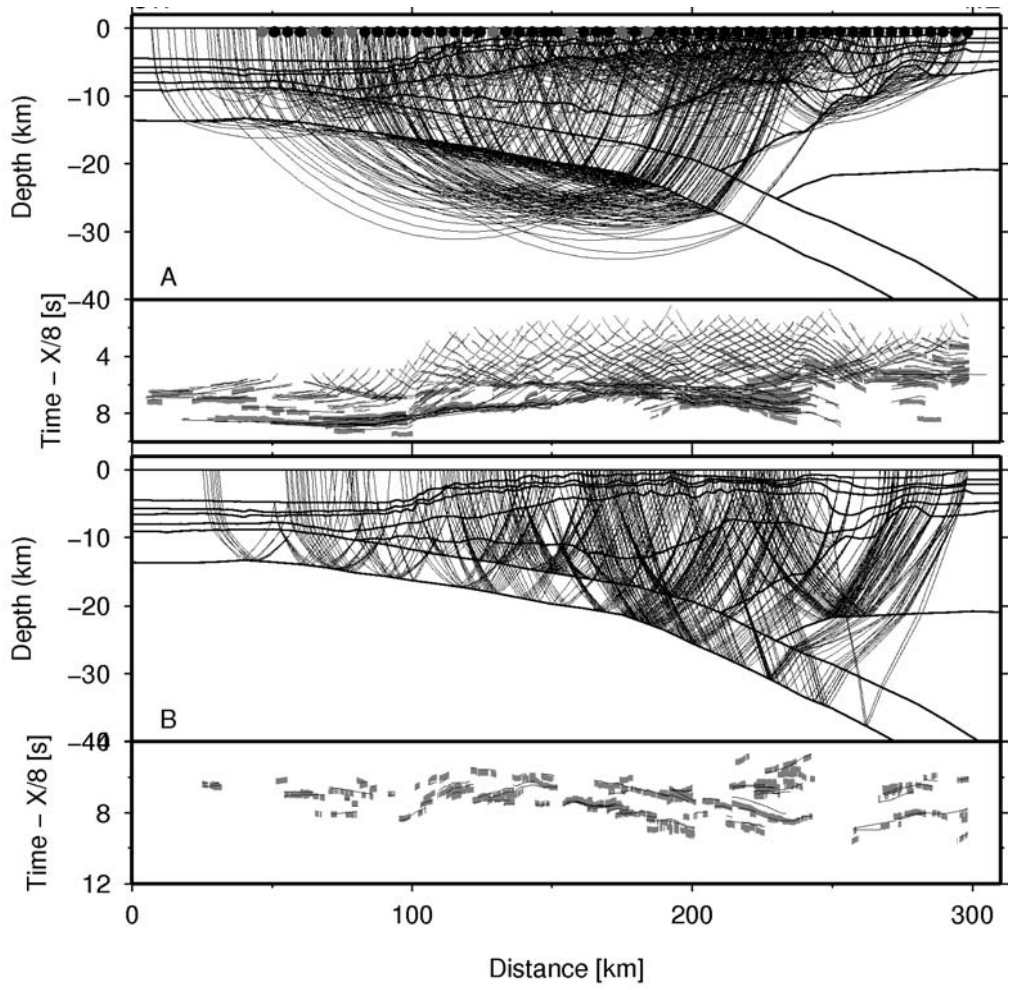

970

Figure 12

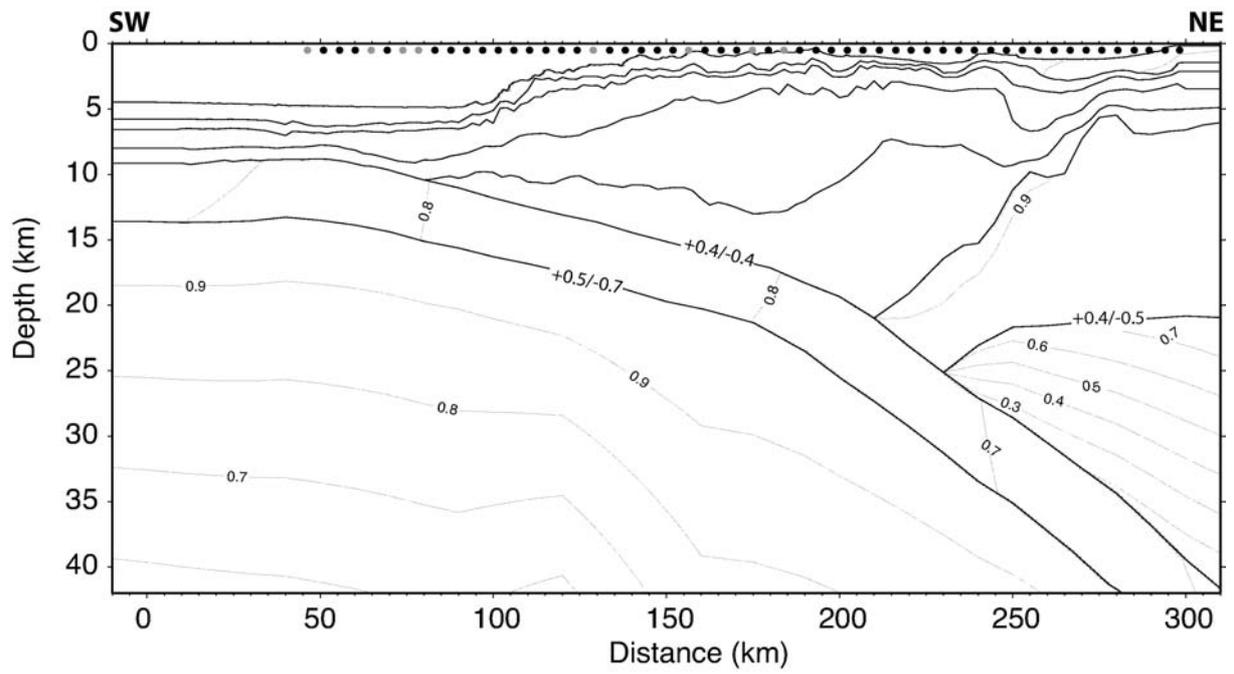

Figure 13

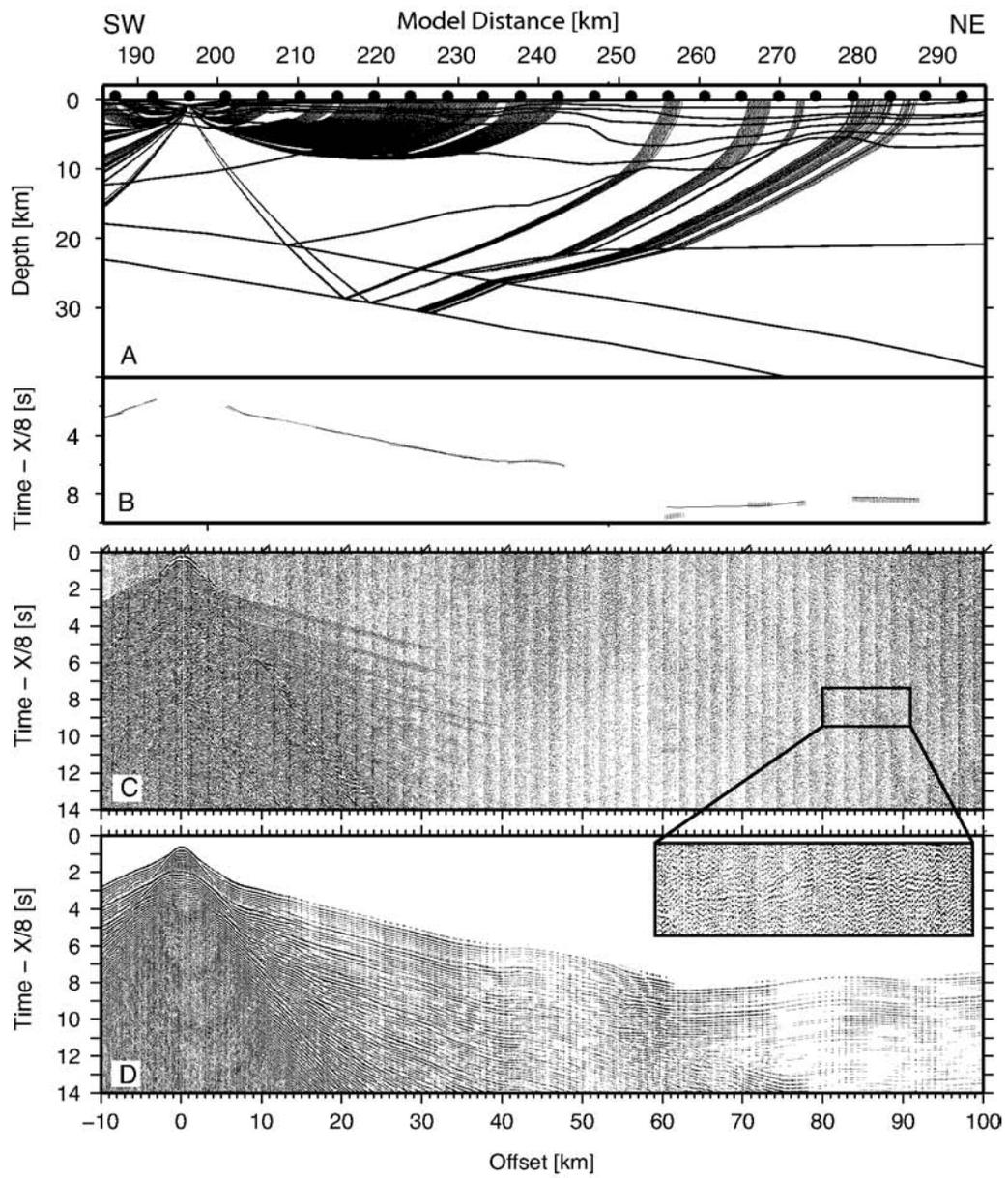

Figure 14

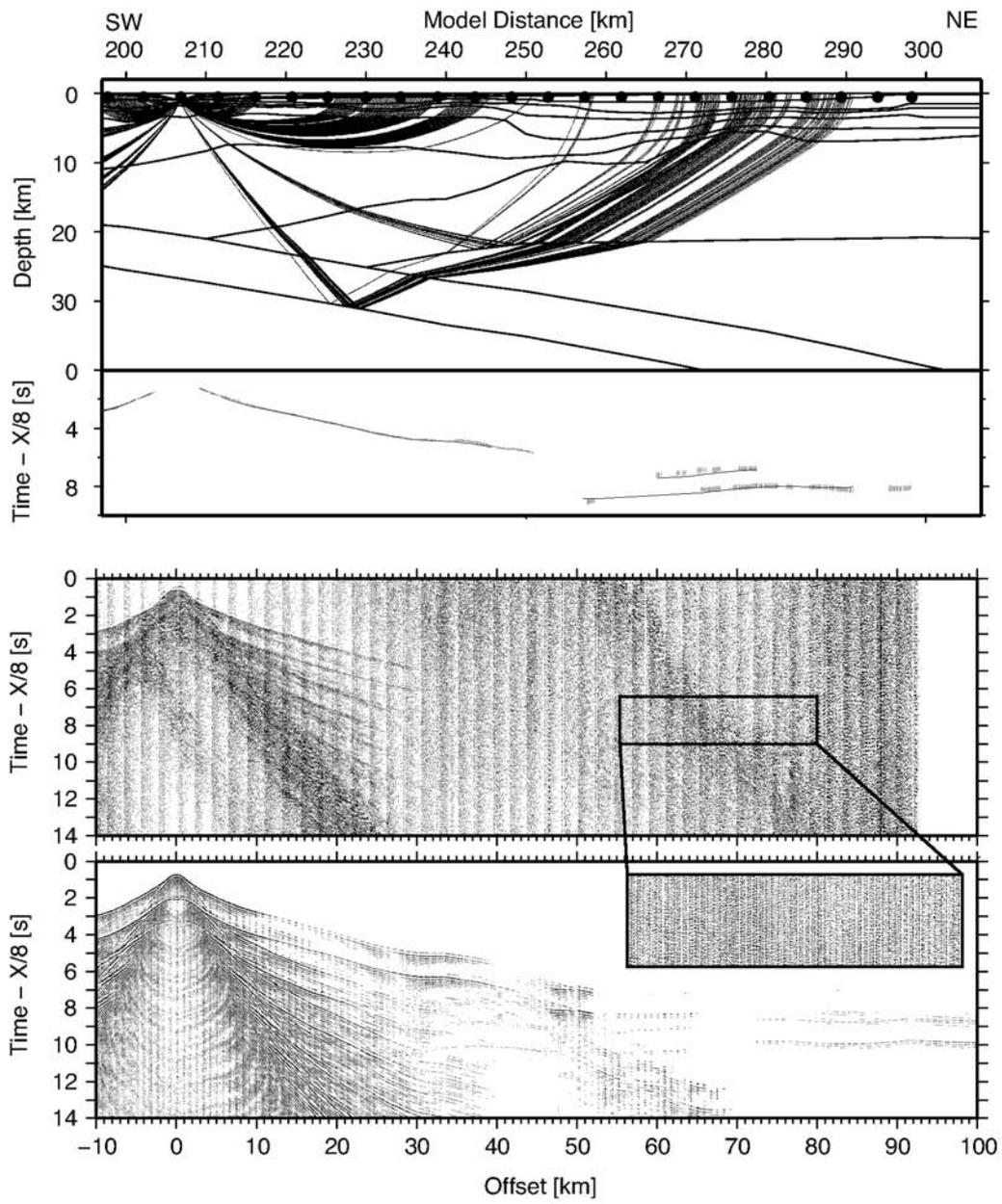

Figure 15

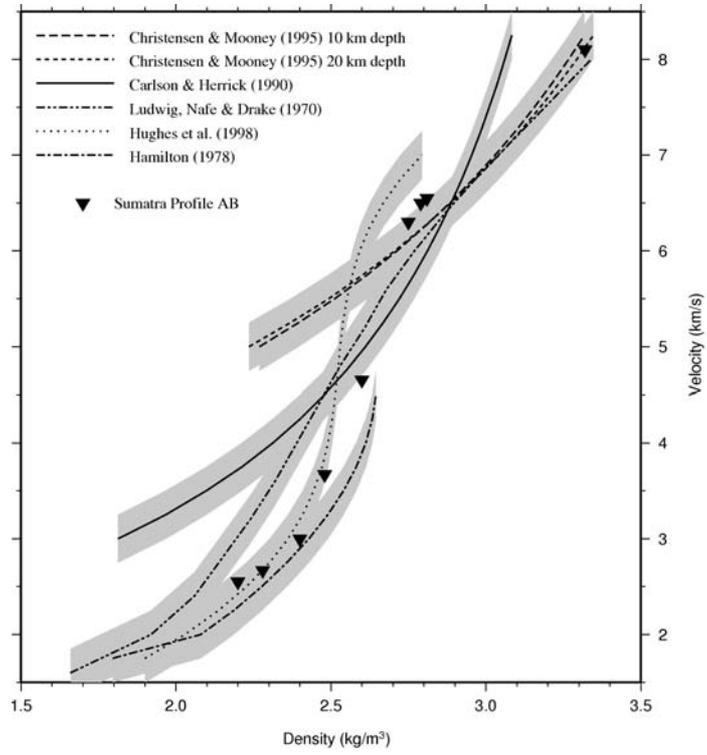

Figure 16

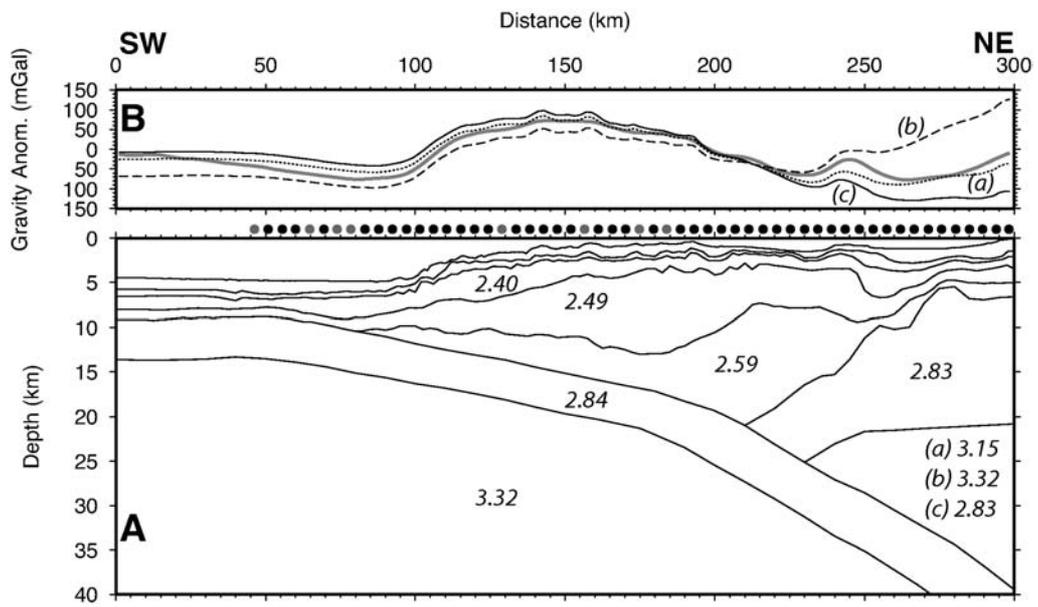

Figure 17

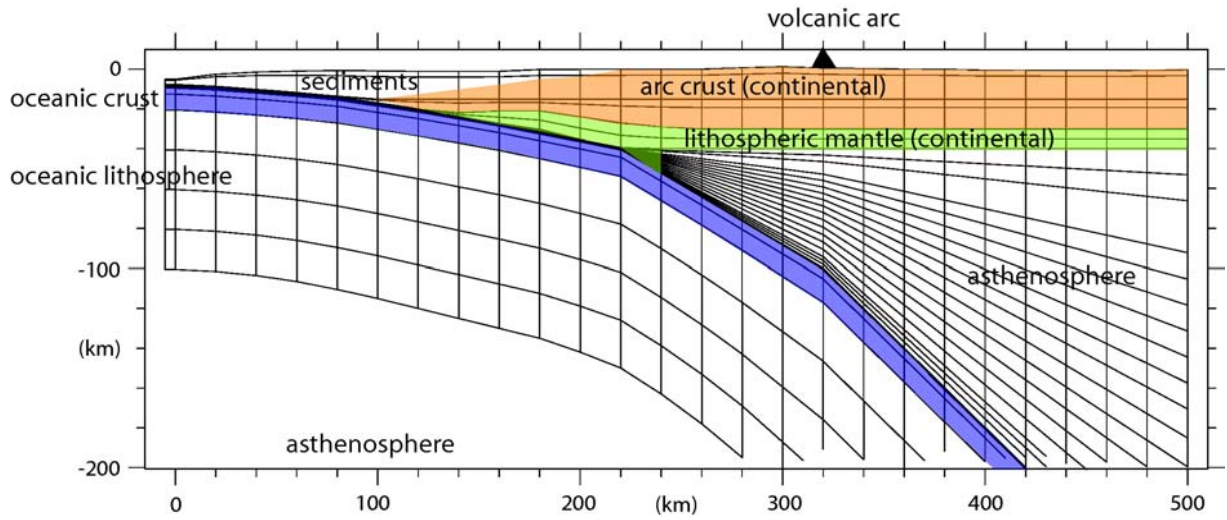

*Figure 18*

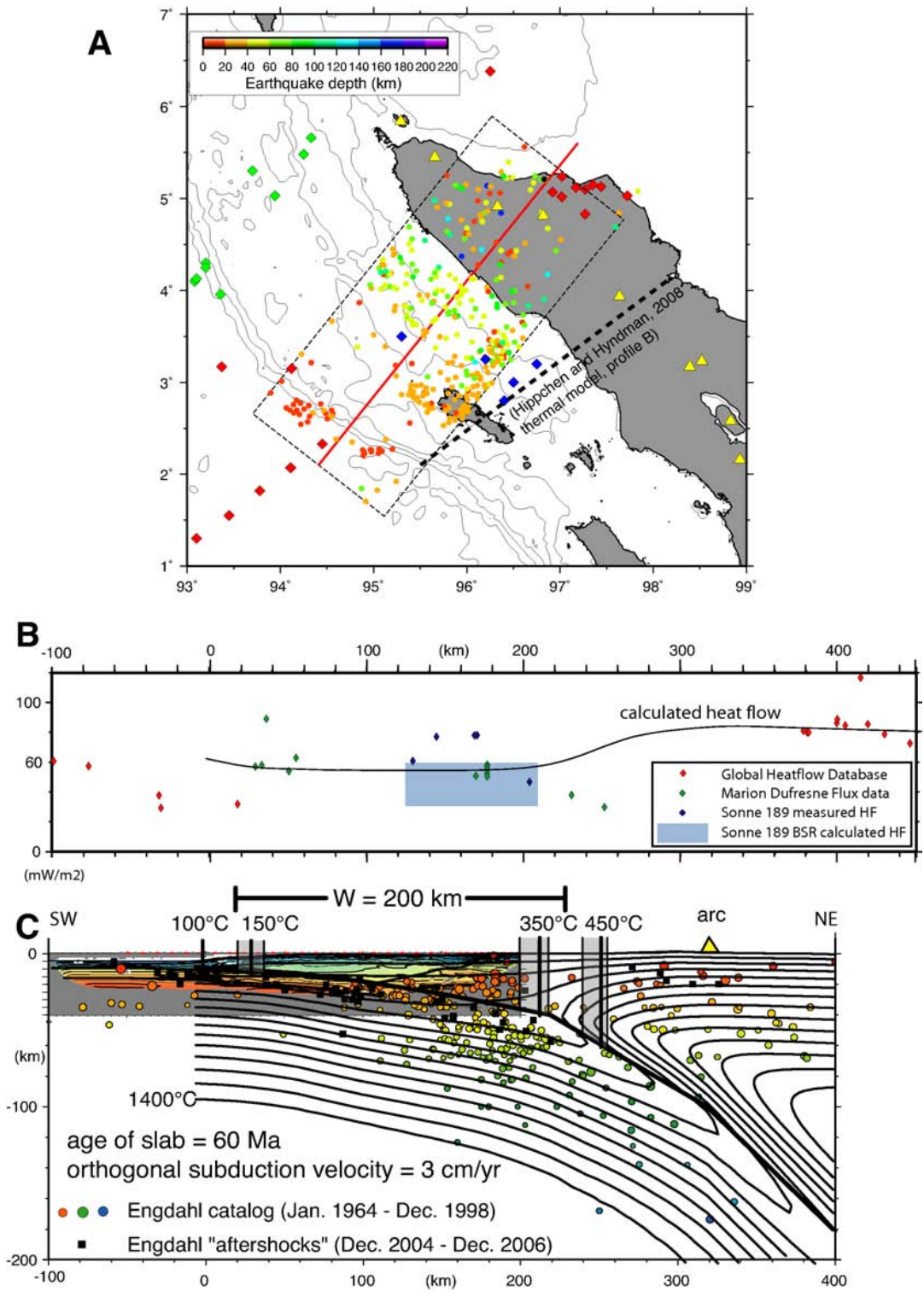

*Figure 19*